# Deterministic reshaping of single-photon spectra using cross-phase modulation

**Nobuyuki Matsuda**

NTT Basic Research Laboratories, NTT Corporation, Atsugi, Kanagawa 243-0198, Japan

m.nobuyuki@lab.ntt.co.jp

The frequency conversion of light has proved to be a crucial technology for communication, spectroscopy, imaging, and signal processing. In the quantum regime, it also offers great potential for realizing quantum networks incorporating disparate physical systems and quantum-enhanced information processing over a large computational space. The frequency conversion of quantum light, such as single photons, has been extensively investigated for the last two decades using all-optical frequency mixing, with the ultimate goal of realizing lossless and noiseless conversion. I demonstrate another route to this target using frequency conversion induced by cross-phase modulation in a dispersion-managed photonic crystal fiber. Owing to the deterministic and all-optical nature of the process, the lossless and low-noise spectral reshaping of a single-photon wave packet in the telecommunication band has been readily achieved with a modulation bandwidth as large as 0.4 THz. I further demonstrate that the scheme is applicable to manipulations of a nonclassical frequency correlation, wave packet interference, and entanglement between two photons. This approach presents a new coherent frequency interface for photons for quantum information processing.

The photon appears to be a natural choice as an information carrier in quantum communication, which facilitates entanglement distribution over long distances. Such quantum networking (1), which offers opportunities for various technologies in the field of quantum information processing (2–4), requires a coherent optical link to be established between physical systems that have disparate properties such as frequency responses. Thus, the capacity to harness the center frequency and shape of a photon spectrum is vital (5). Such technology will also provide direct control of the quantum state of photons encoded in a frequency degree of freedom of light, which naturally serves a wealth of state space for large-scale quantum computation and simulation (6–10).

Since its first experimental demonstration in 1992 (11), the center frequency conversion of nonclassical light has been widely investigated using nonlinear three- or four-wave mixing (FWM) (12–19), where photon frequencies can be converted with mediation provided by the energy of other input pump fields. The intense pump required for highly efficient conversion tends to simultaneously create noise photons, whereas the demand for scalable quantum networking provides strong motivation for research into the simultaneous realization of lossless and noiseless conversion; for example, an internal efficiency of more than 80% with a signal-to-noise ratio of over 100 was demonstrated for a single-photon state (18). Furthermore, the other key function, spectral reshaping, was recently appended to the frequency conversion processes (20–22), and research aimed at high-efficiency reshaping of single photons is in progress (23, 24). At the same time, the search for other schemes that can harness the spectral property of photons, ideally based on a deterministic process, may offer a new approach toward scalable quantum networking.

Another nonlinear optical interaction that enables frequency conversion is cross-phase modulation (XPM) (25, 26), which is a third-order nonlinear optical effect that enables the phase of an optical field to be controlled by another field. XPM provides a function for spectral reshaping via dynamic phase alteration of an optical field, which is widely applied to the frequency conversion of high-speed optical signals (26), to biological imaging (27), and even to testing the event horizon (28) in the classical regime. The XPM-induced phase shift at the quantum level has been extensively investigated (29, 30) with a view to realizing scalable quantum logic gates (4) and quantum nondemolition measurement (31), thanks to the deterministic nature of the XPM interaction (32). Thus, it also has the potential for realizing highly efficient spectral reshaping that is applicable to single photons. Experiments in the classical regime achieved reshaping with a conversion bandwidth of more than 1 THz (26) as well as key tasks for communication including uniform spectral shifting (33), spectral compression (34), time lensing (35), and pulse retiming (36). However, XPM-based spectral manipulation has not yet been applied to single photons.

Here, I present the first experimental reshaping of spectral distributions of single photons using XPM. The deterministic nature of the process made it possible to realize the spectral conversion of single-photon wave packets without an observable interaction loss caused by the conversion. Using the high bandwidth of the dynamic frequency shift induced by subpicosecond optical pulses, I report the postgeneration conversion of the nonclassical joint spectral correlation of photon pairs. Furthermore, I show that the wave packet interference and entanglement between the reshaped single photon and another photon remain, directly demonstrating the applicability of XPM reshaping to quantum information technology.

A wave packet consisting of a single photon acquires a new frequency component via XPM induced by an intense control pulse propagating at the same speed in an optical fiber as in an optical Kerr medium (Fig. 1). In the classical picture, the presence of a control pulse with an intensity profile $P(t)$ leads to an intensity-dependent variation in the refractive index of the material, which is experienced by the signal field as a phase shift $\phi(t) \propto P(t)$. This gives rise to the instantaneous frequency shift $\Delta\omega(t) = -d\phi(t)/dt$ (25). Unlike frequency mixings, spectral reshaping is not accompanied by the center frequency translation of the modulated photons.

The dynamic frequency shift of single photons has also been demonstrated using electro-optic (EO) modulators driven by a radio-frequency electric field (7, 37, 38). Nonetheless, XPM can provide a larger conversion bandwidth [> 1 THz (26)] than that obtained with the EO modulator approach because the control field is an optical pulse that can be made ultrafast in the femtosecond regime. This is useful for manipulating the frequencies of photons generated in a spontaneous process in nonlinear crystals, which have played pivotal roles in a number of quantum information experiments (39–43). Furthermore, $\Delta\omega(t)$ can be controlled simply by tailoring the control pulse envelope function $P(t)$. Hence, the all-optical scheme will offer various spectral reshaping capabilities (33–36) incorporated with well-developed ultrafast pulse shaping technology (44).

- 3 -

The quantum field to be converted now has an intensity that is considerably lower than that of the control pulses. Accordingly, a large frequency separation is needed between the two interacting fields to eliminate potential noise caused by the nonlinear spectral broadening of the control pulse. Moreover, the two interacting fields must propagate at the same pace; otherwise, $\phi(t)$ will become constant over the entire wave packet and cancel the frequency shift.

Photonic crystal fiber (PCF) (45) is a Kerr medium that fulfills this criterion. I used a PCF (NL-5.0-1065 from NKT Photonics A/S, 1 m long) that has the dispersion property shown in Fig. 1B, which I measured using white light cross-correlation interferometry (46). It exhibits two widely separated wavelengths with the same group velocity, thanks to the dispersion property, which is managed by arranging the air-hole claddings that surround the silica core (45). Such a dispersion engineering property, unachievable by standard optical fibers, is widely used in various applications including supercontinuum generation (45), FWM-based quantum frequency conversion (15), and test beds for single-photon nonlinearity (30) and the event horizon (28).

Figure 1C shows a schematic illustration of the experimental setup. Here, I reshape the frequency of one of the correlated photons (labeled signal and idler) created via type II spontaneous parametric down-conversion (SPDC) in a periodically poled potassium titanyl phosphate (PPKTP) crystal (41). The center wavelength of the control field $\lambda_c$ was chosen to be 756 nm, which satisfies $v_g(\lambda_c) \sim v_g(2\lambda_c)$, where $v_g$ is the group velocity of the PCF. In this way, both the XPM control pulses and SPDC excitation pulses can be fed from a pulsed laser source (see Materials and Methods for details).

In Fig. 2, I demonstrate the reshaping of the frequency correlation by plotting the measured joint spectral intensity (JSI) $|S(\omega_s,\omega_i)|^2$ of the photon pairs for various time delays, $\Delta T$, of the signal photon with respect to the control pulses. Here, $\omega_{s(i)}$ is the angular frequency of the signal (idler) photons and $S(\omega_s,\omega_i)$ is the joint spectral amplitude (JSA) of a biphoton whose state is described by $|\psi\rangle \propto \int d\omega_s d\omega_i S(\omega_s,\omega_i) \hat{a}_s^\dagger(\omega_s) \hat{a}_i^\dagger(\omega_i) |0\rangle$, where $\hat{a}_{s(i)}^\dagger(\omega_{s(i)})$ is the creation operator of a photon in the signal (idler) mode (40). Because there is no XPM for $\Delta T$ values much larger than the temporal width of the control pulses (0.78 ps), the JSI in Fig. 2A is identical to the initial JSI of the photon pairs emitted from the source (see fig. S1).

At $\Delta T$ = 0.83 ps, where the signal photons are mainly synchronous at the trailing edge of the control pulses in the PCF, the JSI is entirely blue-shifted along the axis of the signal wavelength (Fig. 2B). Here, the frequency shift is approximately 3.2 nm (0.4 THz). This shows that the frequency correlation between photons is transferred to new frequency sets. When the photons stay at around the peak position of the control pulse in the PCF ($\Delta T$ = 0.37 ps, Fig. 2C), the JSI is broadened along the horizontal axis, revealing a further reshaping capability. Control of the JSI (JSA) is an essential task in quantum information science and has been developed by engineering the characteristics of nonlinear crystals or pump pulses for photon pair generation (40, 47). The result constitutes the first experimental observation of postgeneration conversion over the JSI via a frequency conversion scheme.

Figure 3A shows the $\Delta T$ dependence of the signal photon spectra heralded by the detection of idler photons. Here, the center wavelength of the band-pass filter for the idler photons [tunable band-pass filter 2 (TBPF2)] is fixed at $2\lambda_c$ (1512 nm) to make it possible to observe the signal photon component that was originally correlated with the idler photons at that wavelength. The heralded spectrum is markedly modified as $\Delta T$ varies. For instance, as expected (25), most of the spectrum is red- or blue-shifted when photons are at the trailing or rising edge of the control pulses, respectively. Figure 3B is a plot of the sum of the coincidence counts for each $\Delta T$. The total coincidence count remains constant regardless of the delay, demonstrating that the conversion successfully occurred without an observable photon loss at least for a data fluctuation of 2.2% (SD). A numerical simulation of the frequency modulation is performed on the basis of coupled nonlinear Schrödinger equations including XPM, self-phase modulation, and dispersion (see section SI). The simulation result is obtained, as shown in Fig. 3C, with reasonable free-fitting parameters that describe well the characteristics of the spectral shifts seen in the experimental data.

Next, I use the reshaping scheme to tailor a two-photon interference, which lies at the heart of photonic quantum information technologies (1–4). The observation of this two-photon interference with frequency-converted photons is also a preliminary step toward the construction of a quantum network using distinct physical systems and quantum information processing using photons with different colors (13, 18). In the following, the interference is shown to occur even after XPM reshaping. The experimental setup is modified (see Materials and Methods) so that the Hong-Ou-Mandel (HOM) interference (39) will occur at the NPBS if the signal and idler photons are indistinguishable as regards any physical degree of freedom. I start with photon pairs with nondegenerate center wavelengths whose single-count spectra and JSI are shown in Fig. 4A (blue curves) and Fig. 4C (left). For these photons, a two-photon interference fringe, which is the coincidence count rate R versus the signal-to-idler arrival time difference δτ at NPBS (provided by the delay line), is observed as shown in Fig. 4B (blue squares). The poor visibility is due to the small spectral overlap (23%) between the initial photons.

Then, an XPM-induced blue shift is applied to the signal photons. When $\Delta T$ = 0.88 ps, the spectrum of the signal photons is altered into the distribution shown by the red curve in Fig. 4A, which attained an increased spectral overlap of 70% with that of the idler photons. Now, I obtain an interference fringe exhibiting a clear HOM dip displayed as red dots in Fig. 4B. The visibility of the dip, $V = |(R_{classical} - R_{quantum})/ R_{classical}|$, is 0.84 ± 0.01. This exceeds the classical limit of 0.5 and confirms that the coalescence of the photon pairs took place at the NPBS thanks to the successful reduction of biphoton distinguishability. Without the accidental (background) coincidence count, $V$ = 0.87 ± 0.01, which is explained by the upper-bound visibility of 0.91 calculated (see section SII) from the JSI after XPM (Fig. 4C, right). The triangular shape of the dip is Fourier-related to the sinc-shaped JSA (see section SI) of the correlated photons created in SPDC (38, 40), as the small side lobes of the sinc squared function of the JSI can be slightly seen in the measured data. The low background levels in Fig. 4A, from which only detector dark counts have been subtracted, demonstrate that reshaping is accompanied by low noise. The observed background single count induced by the control field is less than 100 Hz per TBPF window (0.4 nm). $R_{classical}$ ($R$ at a sufficiently large $\delta\tau$) is largely maintained regardless of XPM thanks to the lossless conversion. The subtle discrepancy is due to the nonlinear polarization rotation of the signal photons (see section SIII for details).

In a two-photon interference setup, a bump will appear in the coincidence count when biphotons have an antisymmetric (fermionic) wave function in a subspace, such that $S(\omega_s,\omega_i)$ = − $S(\omega_i,\omega_s)$ in frequency, where the symmetry refers to the exchange of particles (42, 48). This is because the entire photon (boson) wave function is inherently symmetric, and thus an antisymmetric wave function introduces antisymmetry into the spatial part of the wave function, which is filtered by the NPBS. This results in antibunching of the photons, which leads to an $R$ greater than $R_{\text{classical}}$, the observation of which is a sufficient criterion for entanglement (42, 48). Such phenomena are also observed by introducing an antisymmetric wave function in other subspaces such as polarization (43). Now, I attempt to observe the phenomenon by increasing the antisymmetric part in a frequency subspace via XPM.

Starting with the photon pairs whose spectra are shown as blue curves in Fig. 4D and the left panel in Fig. 4F, I obtain the two-photon interference fringe shown as blue squares in Fig. 4E. Then, the marginal idler spectrum is reshaped so that the JSA of the photon pairs will gradually acquire antisymmetric components. When the idler photons reach the state with the single and joint spectra indicated by the red solid curve in Fig. 4D and the right panel in Fig. 4F, the two-photon interference fringe turns into that represented by the red dots in Fig. 4E. Now, the fringe shows a bump at around $\delta\tau = 0$. The visibility $V$, obtained using the aforementioned definition, is $V = 0.082 \pm 0.019 > 0$. This is as large as $V = 0.083 \pm 0.018$, which is the maximum value obtained for the source when the antisymmetric component is increased by scanning the PPKTP temperature but without XPM (fig. S4B). The result demonstrates that, even throughout the XPM reshaping, the quantum entanglement of the biphoton remains, rendering the potential for application to quantum information processing.

Reshaping over a marginal spectral distribution of single photons has been demonstrated using XPM. Lossless, low noise, and a large bandwidth nature enabled me both to reshape the spectral correlation of the photons and to tune the nonclassical interference between photons. The amount of the spectral shift (0.4 THz) was limited by the group velocity dispersion (GVD) of the PCF used. The dispersion engineering of PCFs (45) will be useful for reducing the GVD, which will lead to a larger modulation bandwidth (see section SV). A highly developed technology for ultrafast pulse shaping (44) will expand the reshaping functionalities (25, 33–36). Lossless conversion will easily enable the simultaneous reshaping of multiple photons. These features will allow one to shape wave packets of photons for quantum networking and perform the direct manipulation of the quantum states of photons encoded in frequency, where a large computational space is naturally available. The scheme will provide an extension to the existing tools for quantum information processing that requires the coherent manipulation of the frequency property of photons.

**Acknowledgements**

I am grateful to H. Takesue, W. J. Munro, K. Shimizu, K. Naganuma, and A. Ishizawa for the fruitful discussions. This research was supported by a Grant-in-Aid for Young Scientists (no. 26706021) from the Japan Society for the Promotion of Science.

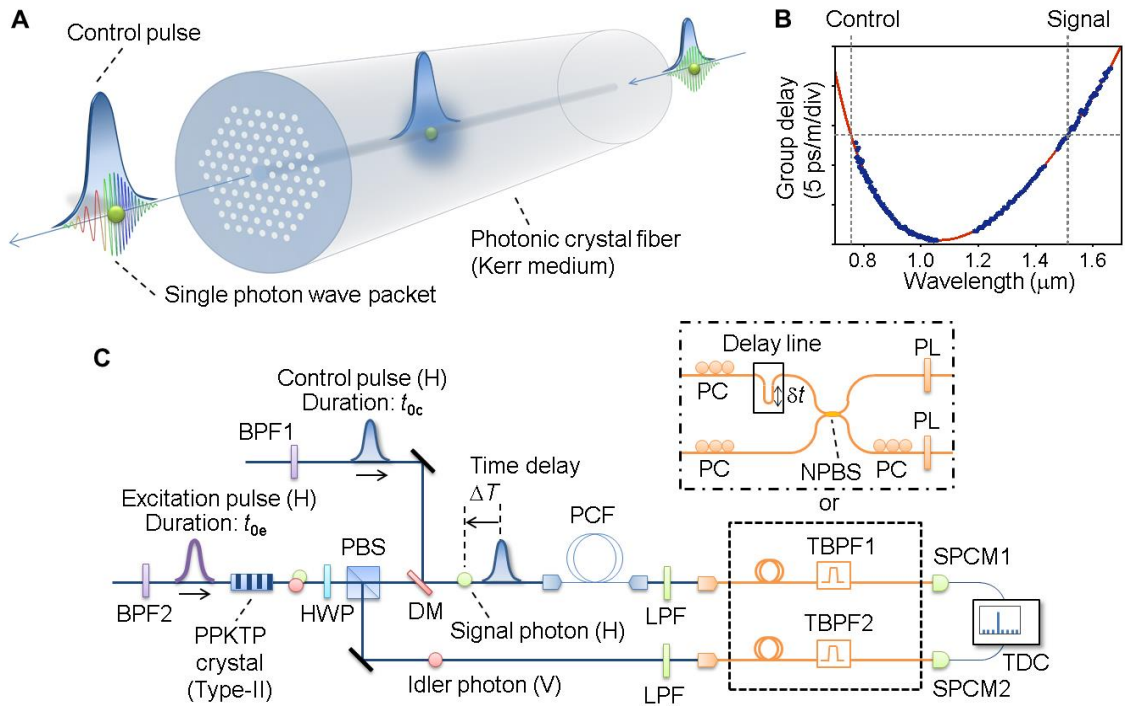

**Fig. 1 Spectral reshaping of a single-photon wave packet using XPM.** (A) Conceptual illustration of the scheme. The control pulse (displayed as an envelope) induces a dynamic nonlinear phase shift, that is, an instantaneous frequency modulation of the single-photon wave packet via XPM in the Kerr medium. (B) Measured group-delay spectrum of the photonic crystal fiber (PCF) used. The solid curve represents a sixth-order polynomial fitting. (C) Schematic of the experimental setup. The control pulse for XPM and the excitation pulse for photon pair generation are obtained from a pulsed laser source. PBS, polarizing beam splitter; DM, dichroic mirror; LPF, long-pass filter; (T)BPF, (wavelength-tunable) band-pass filter; PL, polarizer; PC, fiber polarization controller; NPBS, nonpolarizing 50/50 fiber beam splitter; SPCM, single-photon counting module; TDC, time-to-digital converter. H and V represent horizontal and vertical polarizations, respectively. The paths of signal and idler photons are swapped by the half-wave plate (HWP) in the experiment for entanglement detection (Fig. 4, D to F).

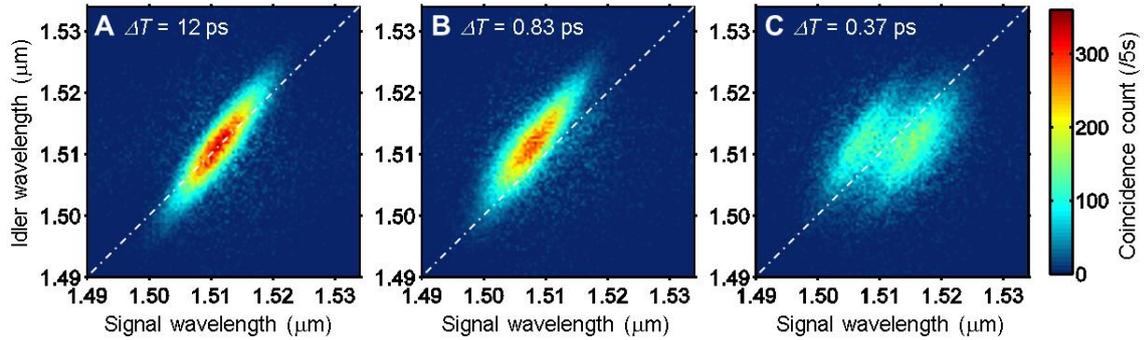

**Fig. 2 Reshaping biphoton frequency correlations.** (A to C) The time delay between the signal photons and the control pulses, ΔT, was set so that the signal photon spectrum was (A) unchanged (original JSA), (B) blue-shifted, and (C) broadened. The zero-detuning lines $\omega_s = \omega_i$ are represented by the dot-dashed lines as a reference.

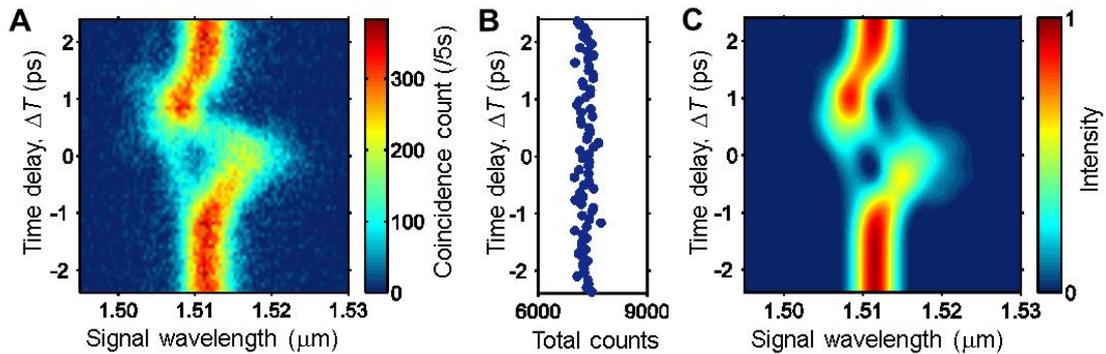

**Fig. 3 Delay dependence of the marginal spectrum of heralded signal photons.** (A) Experimental result. Coincidence counts were recorded as a function of the center wavelength of TBPF1, whereas that of TBPF2 was fixed at 1512 nm. (B) Sum of the coincidence counts for each time delay. The total coincidence count is unchanged regardless of the XPM interaction. (C) Numerical simulation result based on nonlinear-coupled Schrödinger equations (see section SI for details).

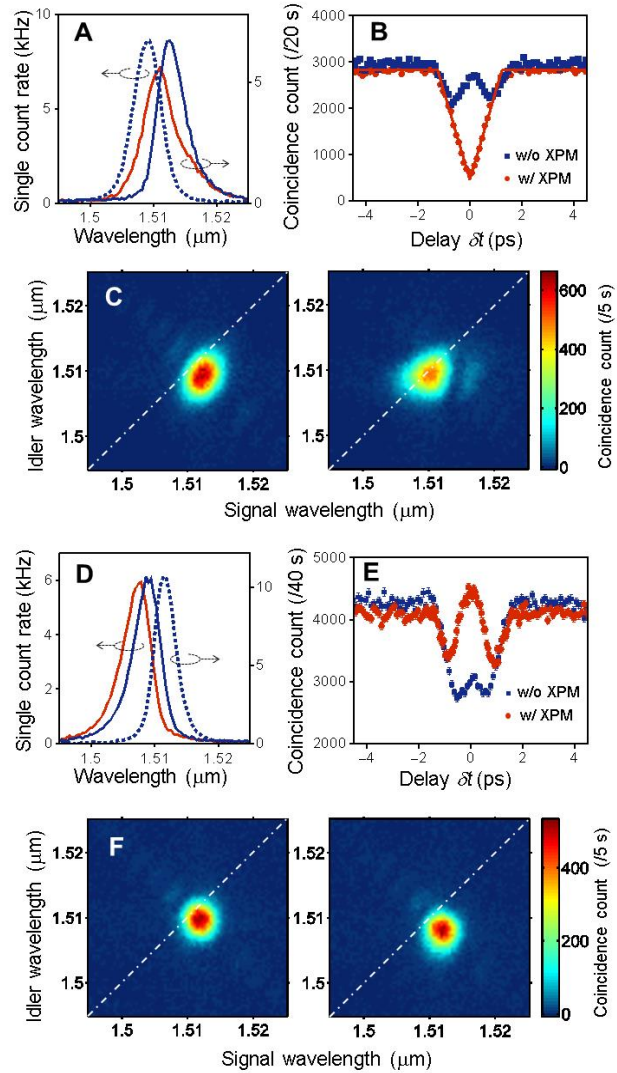

**Fig. 4| Control over nonclassical interference between photons.** (A to C) Engineering biphoton distinguishability in frequency. (A) Single count spectra of signal photons without (blue solid curve) and with (red solid curve) XPM and idler photons (blue dashed curve). (B) Two-photon interference fringes without (blue) and with (red) XPM reshaping. (C) Two-photon JSI without (left) and with (right) XPM. (D to F) Detection of two-photon frequency entanglement after the application of XPM. (D) Single count spectra of signal photons (blue dashed curve) and idler photons without (blue solid curve) and with (red solid curve) XPM. (E) Two-photon interference fringes without (blue) and with (red) XPM, each with different axes for comparison. The bump with XPM reshaping indicates that the entangled component

in JSA acquired an antisymmetric wave function. (F) JSI without (left) and with (right) XPM. In (C) and (F), the zero-detuning lines are provided by the dot-dashed lines as a reference.